# 314-GBaud Single-Wavelength Signaling Generated All-Electronically by a 158-GHz Digital-Band-Interleaved DAC

Di Che[*], Callum Deakin, and Xi Chen

Nokia Bell Labs, Murray Hill, NJ 07974, United States ([*]di.che@nokia-bell-labs.com)

**Abstract** *We demonstrate an all-electronic generation of 314-GBaud PS 8-ASK signals by a 158-GHz digital-band-interleaved DAC. The signals drive a thin-film LiNbO$_3$ modulator that achieves over 300 GBaud single-wavelength Nyquist signaling without optical equalization.* ©2024 The Author(s)

## Introduction

The speed evolution of commercial optical interface has been driven by the ever-increasing symbol rate per wavelength [1]. While a higher symbol rate can be assembled with optical time division multiplexing (OTDM) [2,3] or optical frequency synthesis [4,5], all-electronic high-symbol-rate generation is usually preferred because it requires only one optical carrier and one electrical-to-optical (E/O) modulator, therefore saving cost as well as space. In Fig. 1, we summarize the electronically generated Nyquist symbol rates in recent high-speed optical transmission experiments [6-20]. Note the net data rates (NDRs) per dimension in the figure are either from a 1-D modulation experiment (*e.g.*, intensity-modulation direct-detection, IM-DD) or a 4-D dual-polarization coherent transmission whose NDR is normalized to one dimension. Despite the improvement of NDRs over the past few years, the symbol rates have been limited to a level of 200 GBaud in research labs. Though faster-than-Nyquist (FTN) signaling can push the nominal symbol rate well beyond 200 GBaud (*e.g.*, in [21,22]), it is out of the scope of this paper. The transceiver industry has been actively developing 200-GBaud class techniques to support the next-generation coherent optics at 1.6-Tb/s class. For example, OIF (Optical Internetworking Forum) has announced the 1600ZR project in 2023, aiming to define a multi-vendor interoperable 1.6T coherent interface with a symbol rate requirement beyond 240 GBaud. It is urgently needed to investigate higher symbol rate systems at the next level, to understand the challenge of components/subsystem designs for the generation after.

In this paper, we report single-wavelength Nyquist signaling at 314 GBaud based on a newly designed digital band interleaved (DBI) digital-to-analog converter (DAC) with an analog bandwidth of 158 GHz. We modulate probabilistically shaped (PS) 8-ary amplitude shift keying (8-ASK) with entropies up to 2.9 bits/symbol, achieving an NDR per dimension of 669 Gb/s.

## The 158-GHz DBI-DAC

We follow the same concept of frequency-domain multiplexing as in [23] that achieved up to 114 GHz analog bandwidth previously [20]. We redesign the electronic architecture to extend the DBI bandwidth to 158 GHz. The DBI-DAC consists of three bands as shown in Fig. 2, where the 1st band (low frequency, L-band) covers a frequency range from DC to 82 GHz, the 2nd band (medium frequency, M-band) covers 82 to 116 GHz, and the 3rd band (high frequency, H-band) covers 116 to 158 GHz. The three baseband signals are generated by a 256-GSa/s arbitrary waveform generator (AWG, Keysight M8199B) offering an analog bandwidth of about 80 GHz. We use offline digital signal processing (DSP) to pre-process an arbitrary wideband signal as three spectral partitions to be uploaded to three AWG channels. The signal is first divided into three sub-bands by rectangular digital filters with passbands of [0,82], [82,116], and [116,158] GHz, respectively. The L-band remains at the baseband. The M-band signal is digitally shifted to an intermediate frequency range of [2,36] GHz, and then upconverted by a radio-frequency (RF) local oscillator (LO) at 80 GHz. This leads to an upper sideband (USB) that covers [82,116] GHz as expected, together with an undesired lower sideband (LSB) image. Both the image and the residual 80-GHz LO are rejected by a sharp high-pass filter (HPF), which provides >30 dB suppression within the 2-GHz gap between the M-band signal and the LO. The L- and M- bands (L+M) are then combined via a passive power combiner. The H-band is digitally shifted to an IF range of [2,44] GHz, and then up-converted by an RF LO at 160 GHz. A bandpass filter (BPF) follows the upconverter to select the LSB that covers [116,158] GHz and rejects the residual 160-GHz LO and the undesired USB. This BPF provides >25-dB suppression ratio within the 2-GHz gap between the H-band signal and the LO. Finally, the L+M and the H band signals are combined by a diplexer whose crossover frequency is about 116 GHz, to form a continuous spectrum from DC to 158 GHz. We calibrate the frequency response of RF components and the path delay mismatch and pre-compensate them by offline DSP at the transmitter. We use a 128-GSa/s DAC (Micram DAC5) to generate a seed RF LO at 40 GHz and then use frequency multipliers to generate the desired LO frequencies (2× for the 80-GHz LO, and 4× for the 160-GHz LO). To obtain synchronized RF LOs with respect to the baseband signals, the Micram DAC5 and the Keysight AWG are driven by the same 64-GHz clock source (Keysight M8008A Clock Generator).

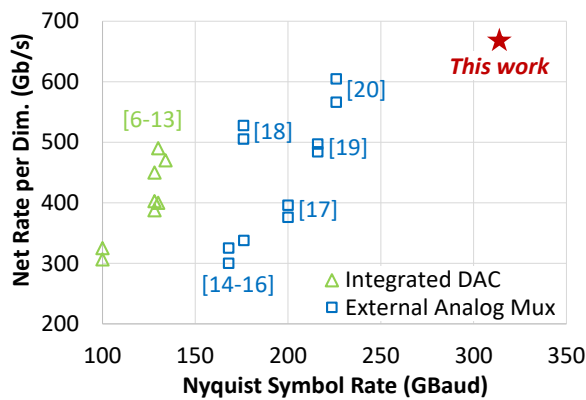

**Fig. 1:** Record Nyquist symbol rates and their NDRs per modulation dimension (Dim.). Mux: multiplexing.

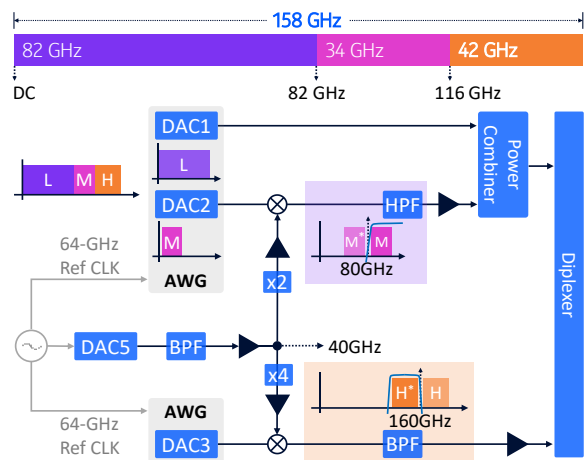

**Fig. 2:** Schematic diagram of the 158-GHz DBI-DAC. HPF: high-pass filter; BPF: band-pass filter.

## Optical Modulation and Detection

We perform a single-wavelength 314-GBaud optical modulation and detection experiment using the setup shown in Fig. 3. The light source is an external cavity laser (ECL) operated at 1550.12 nm with 1-kHz linewidth. The continuous-wave (CW) light is amplified to 27 dBm by an Erbium-doped fiber amplifier (EDFA) and then sent to a thin-film $LiNbO_3$ Mach-Zehnder modulator. The modulator has a 6-dB bandwidth of >100 GHz, and an insertion loss of about 10 dB which is dominated by its grating coupled optical fiber I/O. The modulator is biased near its pull point, leading to amplitude shift keying (ASK) signaling with a 1-D modulation. With 27-dBm optical input, the output power of the amplitude-modulated light is -14.7 dBm. The drive signal generated from the DBI-DAC is added by an RF probe integrated with the diplexer. The modulated optical signal measured on an optical spectrum analyzer (OSA) is shown as an inset in Fig. 3. The spectrum is taken directly from the modulator output without any optical equalization. As seen, the optical spectrum has a width of ~316 GHz with a flat top. As the baseband signal exceeds the bandwidth limitations of our current RF power meter, we can only roughly estimate the peak-to-peak voltage ($V_{pp}$) of the 314-GBd electrical signal from (i) the specifications of the associated analog components, (ii) the $V_\pi$ of the modulator; and (iii) the extrapolation of S21 curve of the modulator. We estimate the driving $V_{pp}$ to be ~1 V.

The bandwidth of the 314-GBaud signal has exceeded the limit of state-of-the-art opto-electronic receivers. Ideally, a reverse version of DBI-DAC (i.e., analog-to-digital converter, ADC) can be built to detect the signal, namely, a >157-GHz photodiode (PD) detects the optical signal, followed by analog band separation and down-conversion to divide the wideband RF output (from the PD) for parallel digitization with multiple lower bandwidth ADCs. However, due to the lack of a high-bandwidth PD as well as additional RF components, we use the optical arbitrary waveform measurement (OAWM) technique [24] instead, that coherently aggregates spectral slices detected by multiple narrow-band optical coherent receivers to reconstruct a wideband signal. As depicted in Fig. 3, our OAWM receiver consists of two coherent receivers, each having a 90° optical hybrid and two 100-GHz balanced PDs. The 314-GBaud signal is divided (from the middle of the spectrum) by a wavelength selective switch (WSS #1) into two spectral slices to be detected by the two receivers. The two LOs of coherent receivers should be phase-locked to recover a phase-coherent pair of spectral slices. Therefore, we pick the LO pair from an optical frequency comb generated by driving a phase modulator (PM) with a high-power RF tone [25]. The seed laser of the comb is a free-running ECL similar to the transmitter one operated around 1550.12 nm. While the intrinsic free-spectral resolution (FSR) of this E/O comb is determined by the RF of 40 GHz, we select two tones spacing at four times of the FSR (*i.e.*, 160 GHz) by a second WSS (#2) and route them to the two coherent receivers. The frequency of each tone is close to the spectral center of a spectral slice for intradyne detection. The PD outputs are sampled by a 4-channel 113-GHz real-time oscilloscope (RTO) sampling at 256 GSa/s. The middle inset of Fig. 3 shows an exemplary pair of digital spectra after frequency offset compensation per slice and amplitude/phase matching between slices, which are ready for spectral stitching in the OAWM DSP.

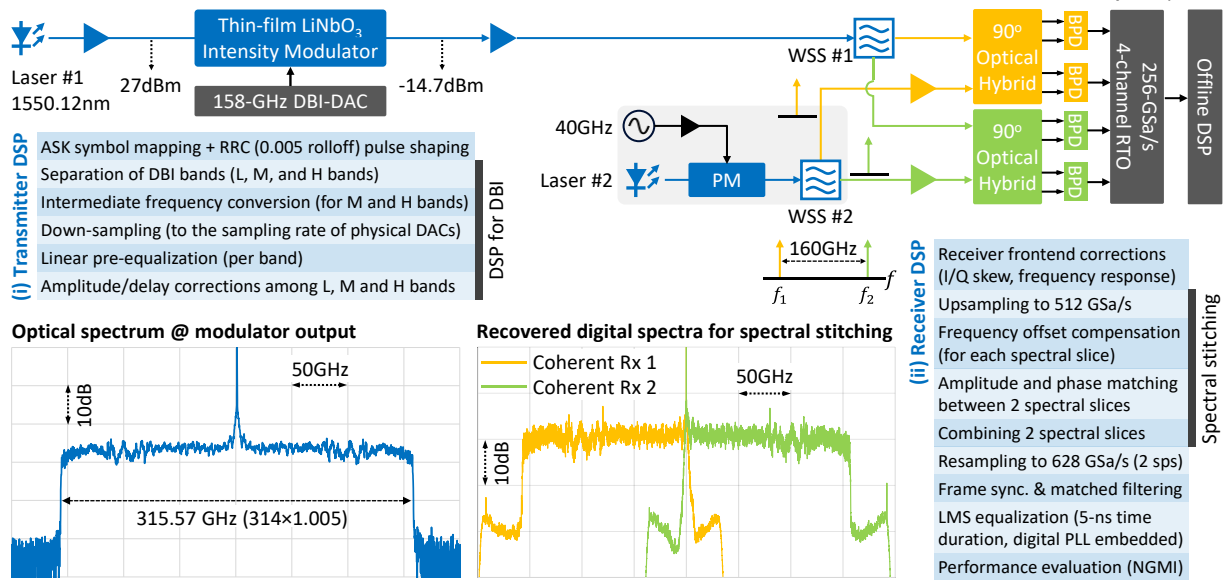

**Fig. 3:** Experimental setup for the optical modulation and OAWM coherent detection of single-wavelength 314-GBaud signals.

**Results and Discussion**

We generate 314-GBaud ASK signals with root-raised cosine (RRC) filtering using a roll-off factor of 0.005 (leading to an optical bandwidth of 314× 1.005=315.57 GHz). The signals then go through the DBI pre-processing DSP as listed in Fig. 3(i). The receiver DSP mainly contains two stages as detailed in Fig. 3(ii). In stage 1, the entire >300-GHz spectrum is reconstructed by OAWM spectral stitching, namely, coherently combining the two spectral slices from the two coherent receivers with precise amplitude and phase matching. In stage 2, 314-GBaud signals are processed by routine coherent DSP at 2 samples per symbol (sps). We choose a rate-0.7436 forward error correction (FEC) code that concatenates a spatially-coupled low-density parity-check (SC-LDPC) inner code (rate 3/4) and a hard-decision BCH outer code (rate 0.9915) [26], with an NGMI threshold of 0.8105. The NDR is calculated as

$$NDR = B \cdot (H - (1-c) \cdot \log_2 M)$$

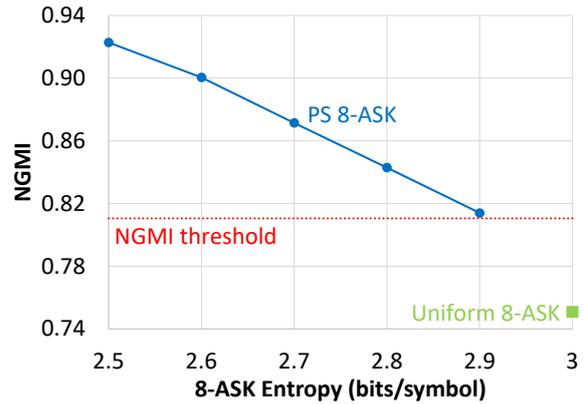

Fig. 4: NGMI as a function of (PS) 8-ASK entropies.

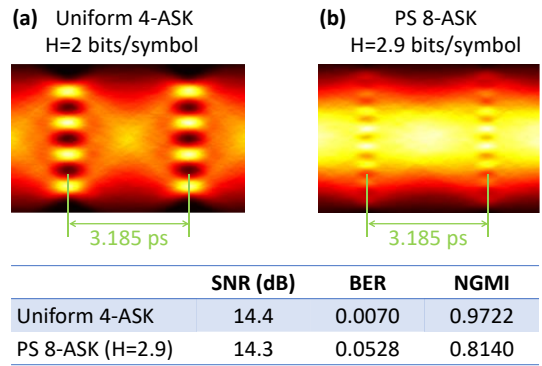

| | SNR (dB) | BER | NGMI |
|---|---|---|---|
| Uniform 4-ASK | 14.4 | 0.0070 | 0.9722 |
| PS 8-ASK (H=2.9) | 14.3 | 0.0528 | 0.8140 |

Fig. 5: Recovered eye-diagrams of 314-GBaud (a) uniform 4-ASK, and (b) PS 8-ASK (entropy of 2.9 bits/symbol) signals.

where $B$ is the symbol rate, $H$ is the entropy (unit of bits/symbol), $c$ is the FEC code rate and $M$ is the modulation order.

In Fig. 4, we vary the entropy of 8-ASK signals and measure their NGMI. PS gains can be maintained in this transmitter-limited system since the DBI pre-processing strongly enhances the peak-to-average power ratio (PAPR), making the system close to an average-power-constrained one (see [27] for more explanations). The highest entropy that yields an NGMI higher than the threshold is 2.9 bits/symbol, leading to an NDR of

$$314 \times (2.9 - (1 - 0.7436)\log_2 8) = 669.07 \text{ Gb/s}$$

for this amplitude-only 1-D modulated signal. Potentially, in a dual-polarization coherent system, the NDR of a single wavelength transmitter could reach 2.67 Tb/s using this DBI structure.

We show two exemplary recovered eye diagrams in Fig. 5 for a uniform 4-ASK and a PS 8-ASK signal (2.9-bit/symbol entropy (H)), respectively, which are digitally reconstructed after the least mean square (LMS) equalization. The uniform 4-ASK signal has a slightly higher SNR than the PS 8-ASK signal due to its lower PAPR [27].

**Conclusions**

We demonstrate an all-electronic generation of single-wavelength 314-GBaud Nyquist signals without the need for optical equalization. Using OAWM coherent detection, we achieve a net data rate up to 669 Gb/s per modulation dimension.